%% file: impeuro.tex
\documentstyle[psfig]{europhys} 
\newcommand{\lesssim}{\raisebox{-.7ex}{\,\protect{$\stackrel{<}{\sim}$}\,}}

\input euromacr

\begin{document}
\euro{xx}{x}{xx-xx}{xxxx}
\Date{}
\shorttitle{M. SCATTONI \etal ANOMALOUS IMPURITY EFFECTS
ETC.}
\title{{\Large Anomalous impurity effects in nonadiabatic superconductors}} 
\author{M. Scattoni\inst{1}, C. Grimaldi\inst{1}\inst{2}\footnote{Present
address: \'Ecole 
Polytechnique F\'ed\'erale de Lausanne,
DMT-IPM, CH-1015 Lausanne, Switzerland}, 
and L. Pietronero\inst{1}\inst{2} } 
\institute{\inst{1} Dipartimento di Fisica, Universit\'{a} di Roma I ``La
Sapienza", Piazzale A.  Moro, 2, 00185 Roma, Italy \\
\inst{2} Istituto Nazionale Fisica della Materia, 
Unit\'a di Roma 1, Italy}
\rec{}{}
\pacs{
\Pacs{71}{38$+i$}{Polarons and electron-phonon interactions}
\Pacs{74}{20$Mn$}{Nonconventional mechanisms}
\Pacs{74}{62$Dh$}{Effects of crystal defects, doping and substitution}
}

\maketitle 
\begin{abstract}
We show that, in contrast with the usual electron-phonon
Migdal-Eliashberg theory, the critical temperature $T_c$ 
of an isotropic s-wave nonadiabatic superconductor is 
strongly reduced by the presence of diluted non-magnetic impurities.
Our results suggest that the recently observed $T_c$-suppression
driven by disorder in K$_3$C$_{60}$ 
(Phys. Rev. B {\bf 55}, 3866 (1997)) 
and in Nd$_{2-x}$Ce$_x$CuO$_{4-\delta}$ (Phys. Rev. B {\bf 58}, 8800 (1998)) 
could be explained in terms of a
nonadiabatic electron-phonon coupling. Moreover, we predict
that the isotope effect on $T_c$ has an impurity dependence qualitatively 
different from the one expected for anisotropic superconductors. 

\end{abstract}

High-$T_c$ superconductors are 
narrow band systems with Fermi energies ($E_F$) one or two
orders of magnitudes smaller than those of conventional 
superconductors \cite{uemura}. 
In these materials therefore it appears unavoidable to address
the validity of Migdal's theorem \cite{migdal} applied to the electron-phonon 
({\it el}-{\it ph}) interaction and eventually 
other mediators \cite{schrieffer}. 
In fact, in fullerene compounds, bismuth oxides and high-$T_c$
cuprates, the typical phonon frequencies
$\omega_0$ can be comparable to $E_F$, making the quantity
$\lambda\omega_0/E_F$ no longer negligible also for moderate
values of the {\it el}-{\it ph} coupling constant $\lambda$.
Given this situation, the ordinary Migdal-Eliashberg (ME) theory,
which is based on Migdal's theorem, hardly can appropriately
describe the properties of such materials and a more general
treatment of the problem should be employed.

In general, by moving from the ME regime one could end to qualitatively
different situations depending on the values of $\lambda$
and $\omega_0/E_F$. 
For example, very strong {\it el}-{\it ph}
couplings favour the formation of polarons and eventually of
bi-polarons. This picture is certainly beyond the ME regime,
however, as recently discussed in ref.\cite{chakra}, it 
is implausible that it can be at the basis of the phenomenon of
high-$T_c$ superconductivity.
Actually, one could go beyond ME regime without ending
to polaron (bi-polaron) states by considering quasi-free 
charge carriers ($\lambda \lesssim 1$)
coupled nonadiabatically ($0<\omega_0/E_F <1 $) 
to the lattice vibrations. In the following, to distinguish 
this regime from the ME and the polaronic ones we use 
the concept of nonadiabatic fermions, which we define as 
quasiparticles (weakly) interacting nonadiabatically with phonons. 
In practice, such a regime can be formulated perturbatively
by treating $\lambda\omega_0/E_F$ as the small parameter
of the theory. At the zeroth order,
the theory coincides with the ME limit while for finite
values of $\lambda\omega_0/E_F$ the nonadiabatic fermions
display anomalous behaviors, at the same time being
faraway from the crossover to polarons.
This latter feature is ensured by considering values
of $\lambda$ smaller than the critical coupling $\lambda_c$
(which is of order unity)
of the crossover to the polaronic state.
The reliability of such a perturbative approach is suggested
by the comparison with exact results for the one electron
system \cite{capone} and quantum Monte Carlo calculations for the
many electrons case \cite{free}.

The results obtained by previous studies of the nonadiabatic
regime interest both the superconducting transition and the
normal state properties. For example, the inclusion of the first vertex
corrections beyond Migdal's limit have led to the possible
enhancement of $T_c$ with respect to the 
adiabatic ME case \cite{grima1}. Another interesting
outcome is certainly the nonzero isotope effect of the 
effective mass of the nonadiabatic fermions \cite{grima2}.
The latter result provides a possible interpretation of 
recent measurements on cuprates \cite{zhao}.  
  
In this paper,
we present results on the non-magnetic impurity effects
on the critical temperature $T_c$ and its isotope 
coefficient $\alpha_{T_c}$
for an isotropic s-wave nonadiabatic superconductor.
We show that disorder strongly affect the 
nonadiabatic corrections and change their analytic properties
substantially.
As a result, $T_c$ can be strongly lowered with respect to the
pure case, in contrast therefore with the adiabatic ME theory which,
according to Anderson's theorem \cite{anderson}, predicts an insensitivity
of $T_c$ with respect to the presence of weak disorder for
an isotropic s-wave superconductor.

Our results can be of particular interest in light of
the disorder dependence of $T_c$ recently observed in
the fullerene compound K$_3$C$_{60}$ and in the electron doped 
cuprate Nd$_{2-x}$Ce$_x$CuO$_{4-\delta}$ \cite{watson,woods}.
Both compounds have a well-established s-wave symmetry of the gap,
hence the disorder induced suppression of $T_c$ cannot be
explained in terms of d-wave pairing \cite{radtke}. It is however true that
anisotropies could lead to qualitatively the same effect also if
the gap is nodeless \cite{openov}. Our results therefore provide
an alternative interpretation based solely on the nonadiabatic
regime of the el-ph interaction.

\begin{figure}
\protect
\centerline{\psfig{figure=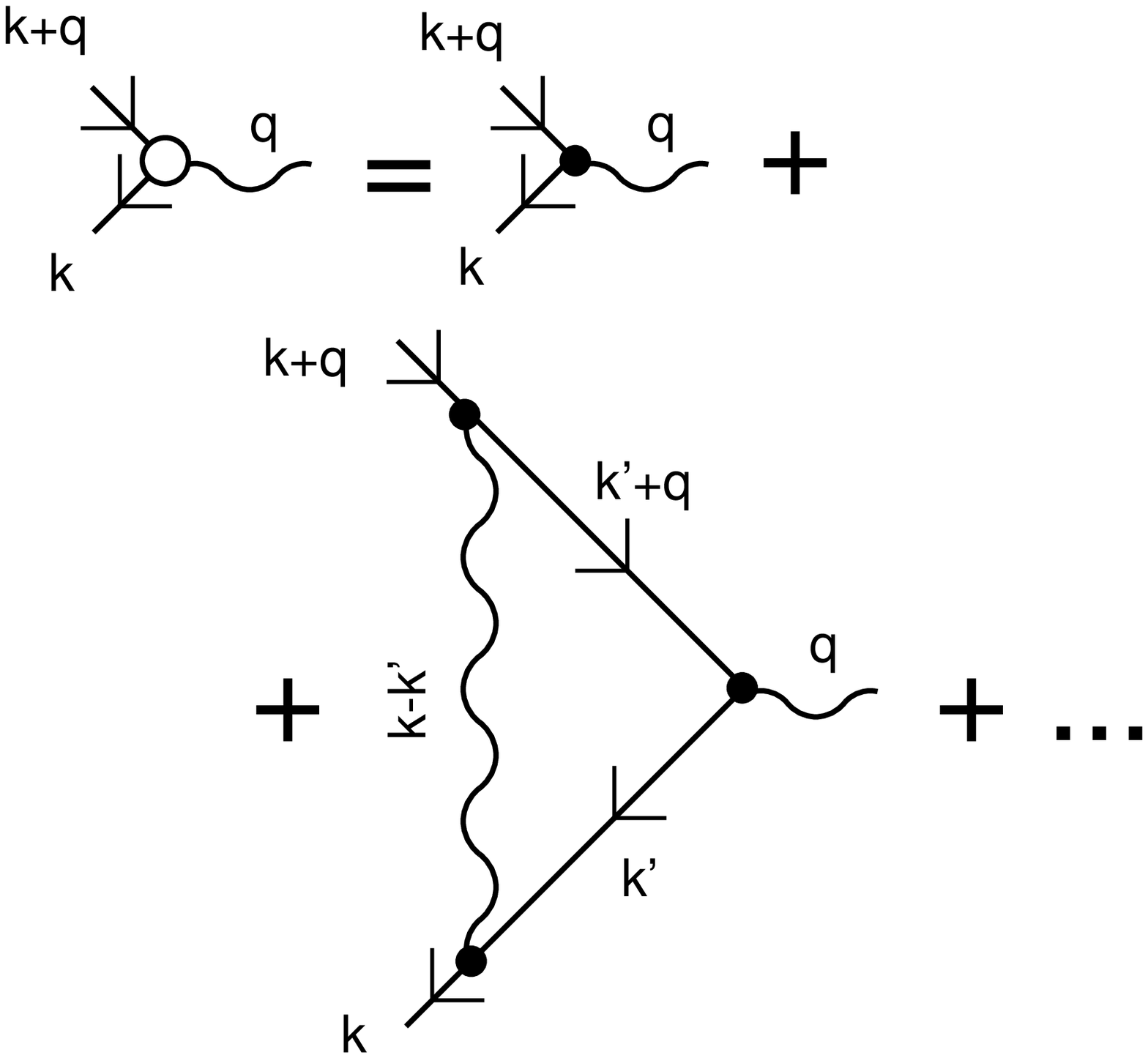,width=5cm}}
\caption{The electron-phonon scattering process.
The filled circles represent $g({\bf q})$.
The last diagram is the first vertex correction
$g({\bf q})P(k+q,k)$ which in the adiabatic limit
gives a negligible contribution according to Migdal's theorem.}
\label{diagr}
\end{figure}

To have a first idea of how disorder affects the
critical temperature of a nonadiabatic superconductor, 
let us consider the first vertex correction in the normal state
depicted in fig.\ref{diagr}.
In the figure, the solid and the wiggled lines are electron and
phonon propagators, $G$ and $D$, respectively, 
while the filled circles represent the
{\it el}-{\it ph} matrix element $g({\bf q})$ for momentum
transfer ${\bf q}$. From the usual rules for diagrams,
the resulting vertex function $P(k+q,k)$ is:

\begin{equation}
\label{vertex1}
P(k+q,k)=\sum_{k'}g^2({\bf k}-{\bf k}')
D(k-k')G(k'+q)G(k') .
\end{equation}
Here, $\sum_{k}$ and $k$ are short notations for
$-T\sum_{\omega_m}\sum_{{\bf k}}$ and ${\bf k},i\omega_m$, respectively.

\begin{figure}
\protect
\centerline{\psfig{figure=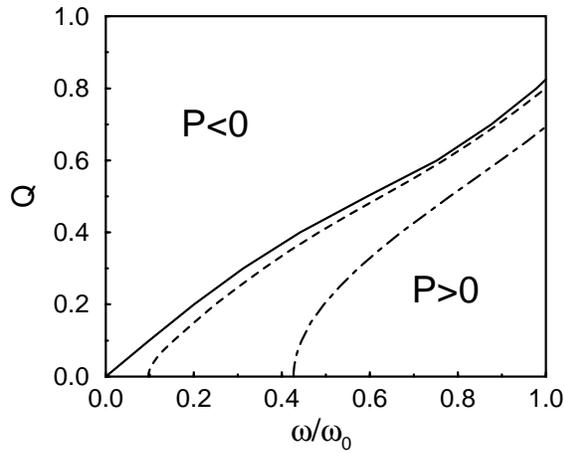,width=11cm}}
\caption{Sign of the vertex function for different values of the
impurity scattering rate $\Gamma$ at $\omega_0/E_F=0.5$. 
Solid line: $\Gamma=0$; dashed line:
$\Gamma=0.1\omega_0$; dot-dashed line: $\Gamma=0.5\omega_0$.}
\label{sign}
\end{figure}

In the presence of impurities, the electron propagators entering 
eq. (\ref{vertex1}) are given by
$G^{-1}(k)=i\omega_m-\epsilon({\bf k})+
i\Gamma\omega_m/|\omega_m|$, where $\epsilon({\bf k})$
is the electronic dispersion and $\Gamma=1/2\tau$ is 
the impurity scattering rate \cite{rick}.  
We neglect for the
moment the self-energy due to the {\it el}-{\it ph} coupling, the effects
of the finite bandwidth on the impurity contribution and the
momentum dependence of $g({\bf k}-{\bf k}')$.
By employing some approximations which will
be specified later, the vertex function (\ref{vertex1}) can be evaluated
as a function of the dimensionless momentum transfer $Q=q/2k_F$,
where $k_F$ is the Fermi momentum, and the exchanged frequency
$\omega$. In fig. \ref{sign} we show the sign of the vertex function
in the $Q$-$\omega$ plane for the half-filling case and for
$\omega_0/E_F=0.5$.
The different lines denote the boundaries where the vertex changes sign.
On the right side of the lines the vertex is positive while 
on the left side is negative.
The effect of impurities is twofold. First, the non-analyticity in 
$\omega=0,Q=0$ found for the pure case \cite{pietro3} (solid line) is removed
when $\Gamma\neq 0$. Second, by increasing the value of $\Gamma$
the boundary lines shift toward higher values of the exchanged frequency,
reducing therefore the region where the vertex is positive.
When we average the vertex over the exchanged momentum and frequency,
we find that by increasing $\Gamma$ the average becomes negative.
Since the vertex correction (\ref{vertex1}) enters into the generalized
Eliashberg equation for a nonadiabatic superconductor \cite{pietro3}, 
we expect from the above result that $T_c$
should be lowered with respect to the pure case.

\begin{figure}
\protect
\centerline{\psfig{figure=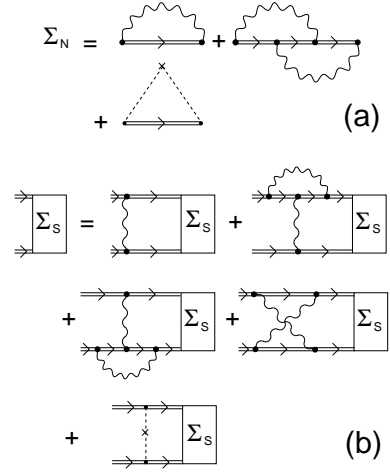,width=5cm}}
\caption{(a): electronic self-energy including the first
vertex correction beyond Migdal's limit and the impurity
contribution.
(b): Self-consistent equation for the anomalous self-energy
generalized to include the first nonadiabatic diagrams (vertex and cross)
and the impurity contribution.} 
\label{self}
\end{figure}

In order to confirm this hypothesis, we solve the 
Eliashberg equations
beyond Migdal's limit by including the effect of disorder in
the Born approximation.
The relevant diagrams for both the normal, $\Sigma_N$,
and anomalous, $\Sigma_S$, self-energies
are depicted in fig.\ref{self}. 
In terms of
the renormalized frequency $W_n=\omega_n Z_n$, where
$Z_n=1-\Sigma_N(i\omega_n)/i\omega_n$, and the renormalized
s-wave gap function $\phi_n=Z(i\omega_n)\Delta(i\omega_n)$,
the generalized Eliashberg equations reduce to:

\begin{eqnarray}
\label{phi}
\phi_n\xi_n  &=&  \pi T_c\!\sum_m\left\{\lambda[1\!+\!2\lambda P(Q_c;n,m)]
D(\omega_n-\omega_m)
+\lambda^2 C(Q_c;n,m)\!\right\}
\!\frac{\phi_m}{|W_m|}\eta_m  ,\\
\label{W}
W_n\xi_n &=& \omega_n\!+\! \pi T_c\!\sum_m\lambda[1\!+\!\lambda
P(Q_c;n,m)]D(\omega_n-\omega_m)
\frac{W_m}{|W_m|}\eta_m ,
\end{eqnarray}
where $\eta_m=(2/\pi)\arctan (E_F/|W_m|)$ is the
finite band-width factor and $\xi_n=1-\eta_n\Gamma/|W_n|$
is the renormalization
induced by the elastic impurities \cite{choi}.

In writing the above equations, we have assumed a dispersionless
phonon spectrum with frequency $\omega_0$ so that
the phonon propagator is given by:
$D(\omega_l)=\omega_0^2/[(i\omega_l)^2-\omega_0^2]$. 
The terms $P(Q_c;n,m)$
and $C(Q_c;n,m)$ are the vertex and cross corrections, 
respectively \cite{grima1,pietro3}.
The parameter $Q_c=q_c/2k_F$ is an upper-cutoff over the momentum transfer 
which follows from the model we use for the {\it el}-{\it ph} coupling
function $g^2(Q)=(g^2/Q_c^2)\theta(Q_c-Q)$ where $\theta$ is the Heaviside
step function \cite{grima1,pietro3}. 
This model simulates the effect of the strong 
electron correlations which lead to mainly forward
{\it el}-{\it ph} scattering processes \cite{zeyher}.
In fact, in a strongly correlated system, a single charge carrier
is surrounded by the corresponding correlation hole whose
linear size $\xi_c$ can extend over many lattice units \cite{dany}. 
The charge carrier can therefore respond only to charge modulations
with wave vector $q_c\lesssim 1/\xi_c$.

The solution of equations (\ref{phi})-(\ref{W}) without
the vertex and cross corrections $P(Q_c;n,m)$ and $C(Q_c;n,m)$
has been already reported by Choi in ref.\cite{choi}. 
The main result was that, for finite values of $E_F$,
$T_c$ can be slightly lowered by the 
presence of impurities.
Here, we solve instead the completely self-consistent equations with
the inclusion of both the vertex and cross corrections.
In evaluating $P(Q_c;n,m)$ and $C(Q_c;n,m)$ we use a density 
of states $N_F$ constant over the entire electron 
bandwidth $2E_F$ and we employ a small momentum transfer approximation
which is valid for small values of the dimensionless cutoff parameter $Q_c$.
From the above discussion, the small $Q_c$ approximation should be
suitable for strongly forward {\it el}-{\it ph} couplings resulting
from the effect of strong electron correlation.
After the integration over the energy and the s-wave average over 
the momentum transfer $Q$ have been performed, 
the vertex and cross corrections reduce to the following form:

\begin{eqnarray}
P(Q_c;n,m)&=&\!\!T_c\sum_l D(\omega_n-\omega_l)
\left\{\frac{}{}B(n,m,l)
+\frac{A(n,m,l)-B(n,m,l)(W_l-W_{l-n+m})^2}
{(2E_FQ_c^2)^2}\right. \nonumber \\
&& \times\left.\left[
\sqrt{1\!+\!\left(\frac{4E_FQ_c^2}{W_l-W_{l-n+m}}\right)^2}
\!-\!1\!-\!\ln\!\left(\frac{1}{2}
\sqrt{1\!+\!\left(\frac{4E_FQ_c^2}{W_l-W_{l-n+m}}\right)^2}
\right)
\right]\right\} ,
\label{vertex2}
\end{eqnarray}

\begin{eqnarray}
C(Q_c;n,m) &=& T_c^2\sum_l D(\omega_n-\omega_l)
D(\omega_l-\omega_m)
\!\left\{2B(n,-m,l)+\arctan\!\left(\frac{4E_FQ_c^2}
{|W_l-W_{l-n-m}|}\right)\right. \nonumber \\
&& \times \left.
\frac{A(n,-m,l)-B(n,-m,l)(W_l-W_{l-n-m})^2}
{2E_FQ_c^2|W_l-W_{l-n-m}|}\right\} ,
\label{cross}
\end{eqnarray}

\begin{equation}
\label{a}
A(n,m,l) = (W_l-W_{l-n+m})\left[\arctan\!\left(\frac{E_F}{W_l}\right)
-\arctan\!\left(\frac{E_F}{W_{l-n+m}}\right)\right] ,
\end{equation}

\begin{equation}
\label{b}
B(n,m,l)  =  (W_l-W_{l-n+m})\frac{E_FW_{l-n+m}}
{[E_F^2+W_{l-n+m}^2]^2}
-\frac{E_F}{E_F^2+W_{l-n+m}^2} .
\end{equation}

\begin{figure}
\protect
\centerline{\psfig{figure=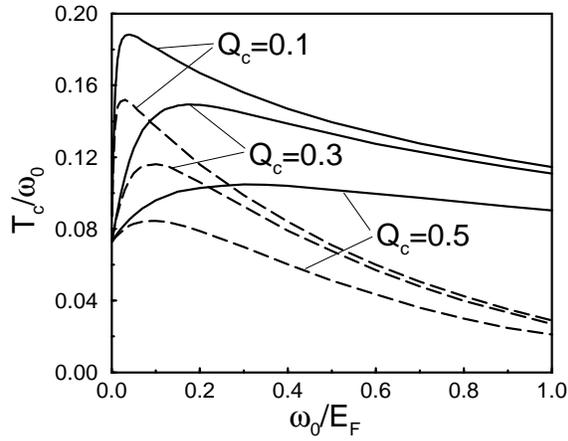,width=11cm}}
\caption{Critical temperature as a function of the adiabatic
parameter $\omega_0/E_F$ for $\lambda=0.7$. 
The solid lines refer to the pure case
($\Gamma=0$) while the dashed lines are the results
for $\Gamma=0.5\omega_0$. Note that at $\omega_0/E_F=0$
the critical temperature is independent of $\Gamma$.} 
\label{Tc_vs_m-g}
\end{figure}

To obtain the critical temperature $T_c$,
we solve numerically the self-consistent 
equations (\ref{phi})-(\ref{b}) for different
values of the adiabatic parameter $\omega_0/E_F$, the momentum
transfer cut-off $Q_c$ and the impurity scattering rate $\Gamma$.
The dependence of $T_c$ upon the adiabatic parameter
is shown in fig.\ref{Tc_vs_m-g} for pure ($\Gamma/\omega_0=0$, solid lines)
and disordered ($\Gamma/\omega_0=0.5$, dashed lines) nonadiabatic 
superconductors for different values of $Q_c$. Our self-consistent
calculations confirm the results of ref.\cite{grima1,pietro3}, 
{\it i.e.}, small values of $Q_c$ lead to an enhancement of 
$T_c$ with respect to the Migdal limit.
Moreover, at $\omega_0/E_F=0$,
$T_c$ is independent of $\Gamma$
in agreement with Anderson's theorem \cite{anderson}. On the other hand,
when $\omega_0/E_F>0$ the impurities lower $T_c$ for all 
values of $Q_c$ and $\omega_0/E_F$. 
We interpret this behavior in terms of the enhanced negative
contribution of the vertex and cross corrections 
induced by the presence of the impurities as depicted in fig.\ref{sign}.
This negative contribution leads to a reduction
of the effective nonadiabatic {\it el}-{\it ph} pairing interaction
resulting in the reduction of $T_c$. 

\begin{figure}
\protect
\centerline{\psfig{figure=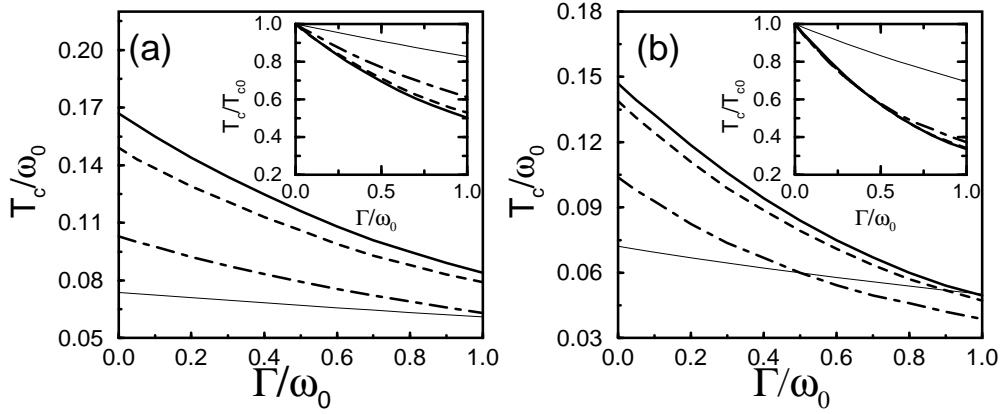,width=13cm}}
\caption{Critical temperature as a function of the impurity
scattering rate $\Gamma$ for $\lambda=0.7$, $\omega/E_F=0.2$ (a)
and $\omega_0/E_F=0.4$ (b). 
The thick (thin) lines refer to the
case with (without) the nonadiabatic corrections (\protect\ref{vertex2})
and (\protect\ref{cross}). Thick solid: $Q_c=0.1$, thick dashed: $Q_c=0.3$,
thick dot-dashed: $Q_c=0.5$.
Inserts: same calculations rescaled with respect to
the critical temperature $T_{c0}$ of pure systems.} 
\label{Tc_vs_G-m}
\end{figure}

In fig.\ref{Tc_vs_G-m} we show $T_c$ as a function of  $\Gamma$ 
for $\omega_0/E_F=0.2$ (a) and $\omega_0/E_F=0.4$ (b).
The thin solid lines refer to the case without vertex and cross corrections
and correspond to the approximation scheme used in ref.\cite{choi}. 
As seen also by the inserts of fig.\ref{Tc_vs_G-m}, when we include the
vertex and cross corrections (thick lines), the reduction of $T_c$ with
the increase of $\Gamma$ can be much
stronger than the reduction given by only the finite bandwidth effects.

So far, we have investigated the effects of disorder on a nonadiabatic
superconductor with an s-wave symmetry of the order parameter.
However, anisotropies of the gap lead to an impurity
dependence of $T_c$ which, to a first approximation, can be described
by using the Abrikosov-Gorkov (AG) scaling law \cite{abrikosov} modified
in order to represent d-wave, anisotropic s-wave and other types of symmetries
of the order parameter \cite{radtke,openov}:
$\ln(T_{c0}/T_c)=\chi[\Psi(1/2+\gamma)-\Psi(1/2)]$,
where $\Psi$ is the digamma function and $\gamma=\Gamma/(2\pi T_c)$.
The parameter $\chi$ is a measure of the anisotropy of the order parameter:
$\chi=1$ ($\chi=0$) for a d-wave (s-wave) superconductor \cite{openov}. 
According to the AG law, for
$\chi\neq 0$  the impurities induce a monotonous 
reduction of $T_c$ in a way qualitatively similar to
the one observed for the nonadiabatic case.

\begin{figure}
\protect
\centerline{\psfig{figure=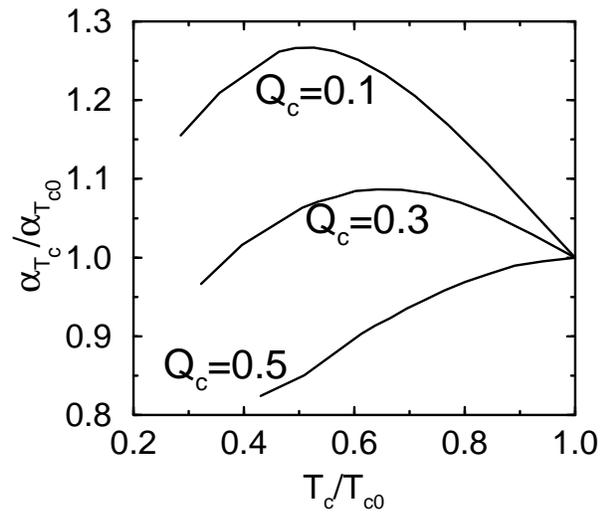,width=12cm}}
\caption{Isotope coefficient $\alpha_{T_c}$ as a function of $T_c$
in the presence of non-magnetic impurities. Both quantities are normalized
to their corresponding values $\alpha_{T_{c0}}$ and $T_{c0}$ 
for the pure limit.
The curves refer to the case of fig. \protect\ref{Tc_vs_G-m}(a).}
\label{alfa_vs_Tc}
\end{figure}

Given the above situation it could be therefore difficult to decide 
whether the $T_c$ suppression
observed in K$_3$C$_{60}$ \cite{watson} and in
Nd$_{2-x}$Ce$_x$CuO$_{4-\delta}$ \cite{woods}
should be ascribed to anisotropies of the order parameter
or instead to the nonadiabatic {\it el}-{\it ph} interaction.
Here, we propose that a more suitable quantity to look at 
could be the ion-mass dependence of the critical temperature.
In fact, the isotope coefficient $\alpha_{T_c}$ resulting from
the AG-type relation is:

\begin{equation}
\label{iso}
\frac{\alpha_{T_c}}{\alpha_{T_{c0}}}=\left[1+\frac{d\ln(T_c/T_{c0})}
{d\ln\gamma}\right]^{-1},
\end{equation}
where $\alpha_{T_{c0}}$ is the isotope coefficient for the pure system.
Since $T_c/T_{c0}$ decreases by increasing $\gamma$,
equation (\ref{iso}) predicts a monotonous impurity induced 
enhancement of $\alpha_{T_c}$ compared with the corresponding 
value in the pure limit \cite{bill}.
Such a behavior is qualitatively different from the one displayed
by an {\it isotropic} s-wave nonadiabatic superconductor. 
In fact, as it is shown in 
fig. \ref{alfa_vs_Tc}, where we report numerical results for $\alpha_{T_c}$
for the same parameters as in fig. \ref{Tc_vs_G-m}(a), 
the decrease of $T_c/T_{c0}$ due to the impurities
is accompanied by a non-monotonous dependence of $\alpha_{T_c}$
at least for small values of the momentum cut-off $Q_c$. Moreover, for larger
values of $Q_c$ the isotope coefficient decreases with the impurity 
concentration showing therefore a behavior opposite to the one 
given by eq.(\ref{iso}).
It would be therefore important to measure the isotope coefficient
and its evolution with the amount of disorder in the two
s-wave superconductors K$_3$C$_{60}$ and Nd$_{2-x}$Ce$_x$CuO$_{4-\delta}$.
Such a measurement could in fact decide
whether the observed $T_c$ suppression in these materials \cite{watson,woods} 
is given by anisotropy or by the
nonadiabatic regime of the {\it el}-{\it ph} interaction.
We note moreover that a measurement of  $\alpha_{T_c}$ {\it vs} $T_c/T_{c0}$
could provide an experimental tool for an estimation of the typical 
momentum scattering $Q_c$ in addition to  the one obtained by tunneling
measurements \cite{umma}.

In summary, we have shown that in the nonadiabatic regime
the critical temperature is lowered by non-magnetic impurities
which also lead to an unusual impurity dependence of
the isotope coefficient.
Our results together with recent measurements on 
K$_3$C$_{60}$ \cite{watson} and 
Nd$_{2-x}$Ce$_x$CuO$_{4-\delta}$ \cite{woods}
suggest that the {\it el}-{\it ph} interaction in these
systems could be in the nonadiabatic regime. We propose also
that in order to confirm or disregard this hypothesis,
a suitable experiment could be the measurement of
the isotope coefficient as a function of the amount of disorder.
We conclude by noticing that, to our knowledge, there are no
experimental results on the effect of disorder on the
bismuth oxides. Since these materials are s-wave 
superconductors \cite{brawner} with estimated 
$\omega_0/E_F\simeq 0.14$ \cite{umma}, we predict that 
in these material $T_c$ should be considerably lowered
by non-magnetic impurities and eventually display an
isotope effect with anomalies as described above.

\stars{C. G. acknowledges the support of a INFM PRA project (PRA-HTCS).}

\newpage

\end{document}

%% file: euromacr.tex
\def\etal{{\hbox{{\tenit\ et al.\/}\tenrm :\ }}}

\def\And{{\rm and\ }}

\def\stars{\bigskip\centerline{***}\medskip}

\newif\ifboo \boofalse

\def\Review#1{\boofalse{\it #1},}
\def\Name#1{{\sc #1},}
\def\Vol#1{\ifboo Vol. {\bf #1}\else{\bf #1}\fi}
\def\Year#1{\ifboo #1\else(#1)\fi}
\def\Book#1{\bootrue{\it #1},}
\def\Page#1{\ifboo {\rm p. #1}\else{\rm #1}\fi}

%% file: impeuro.bbl
\vskip-12pt
\begin{thebibliography}{99}


\bibitem{uemura}
\Name{Uemura Y. J. \etal}
\Review{Phys. Rev. Lett.} \Vol{66} \Year{1991} \Page{2665}.

\bibitem{migdal}
\Name{Migdal A. B.} \Review{Sov. Phys. JETP}
\Vol{7} \Year{1958} \Page{996}.


\bibitem{schrieffer}
\Name{Schrieffer J. R.} \Review{J. Low Temp. Phys.}
\Vol{99} \Year{1995} \Page{377}. 

\bibitem{chakra}
\Name{Chakraverty B. K., Ranninger J., \And Feinberg D.}
\Review{Phys. Rev. Lett.} \Vol{81} \Year{1998} \Page{433}. 


\bibitem{capone}
\Name{Capone M., Ciuchi S., \And Grimaldi C.}
\Review{Europhys. Lett.} \Vol{42} \Year{1998} \Page{523}.


\bibitem{free}
\Name{Freericks J. K., Zlati\'c V., Chung W.,
\And Jarrel M.}
\Review{Phys. Rev. B} \Vol{58} \Year{1998} \Page{11613}.

\bibitem{grima1}
\Name{Grimaldi C., Pietronero L. \And Str\"{a}ssler S.} 
\Name{Phys. Rev. Lett.} \Vol{75} \Year{1995} 
\Page{1158}.

\bibitem{grima2}
\Name{Grimaldi C., Cappelluti E., \And Pietronero L.}
\Review{Europhys. Lett.} \Vol{42} \Year{1998} \Page{667}.


\bibitem{zhao}
\Name{Zhao G. M., Hunt M. B., Keller H. \And M\"{u}ller K. A.}
\Review{Nature} \Vol{385} \Year{1997} \Page{236}. 

\bibitem{anderson}
\Name{Anderson P. W.} \Review{J. Phys. Chem. Solid}
\Vol{11} \Year{1959} \Page{26}.

\bibitem{watson}
\Name{Watson S. K. \etal.}
\Review{Phys. Rev. B} \Vol{55} \Year{1997} \Page{3866}. 


\bibitem{woods}
\Name{Woods S. I. \etal}
\Review{Phys. Rev. B} \Vol{58} \Year{1998} \Page{8800}. 


\bibitem{radtke}
\Name{Radtke R. J. \etal} \Review{Phys. Rev. B} \Vol{48}
\Year{1993} \Page{653}. 

\bibitem{openov}
\Name{Openov L. A.}
\Review{JETP Lett.} \Vol{66} \Year{1997} \Page{661}. 

\bibitem{rick}
\Name{Rickayzen G.}
\Book{Green's Functions and Condensed
Matter} (Academic, New York) \Year{1980}.

\bibitem{pietro3}
\Name{Pietronero L., Str\"{a}ssler S. \And Grimaldi C.}
\Review{Phys. Rev. B} \Vol{52} \Year{1995} \Page{1995}; 
\Name{Grimaldi C., Pietronero L. \And Str\"{a}ssler S.}
\Review{Phys.Rev. B} 
\Vol{52} \Year{1995} \Page{10530}. 


\bibitem{zeyher}
\Name{Zeyher R. \And Kuli\'c M.} 
\Review{Phys. Rev. B} \Vol{54}
\Year{1996} \Page{8985}; 
\Name{Grilli M. \And Castellani C.}
\Review{Phys. Rev. B} \Vol{50} \Year{1994} \Page{16880};
\Name{Keller J., Leal C. E., \And Forsthofer F.}
\Review{Physica C} \Vol{206-207} \Year{1995} \Page{739}.

\bibitem{dany}
\Name{Danylenko O. V. \etal} cond-mat/9710234, Preprint 1997.


\bibitem{choi}
\Name{Choi H. Y.} \Review{Phys. Rev. B} 
\Vol{53} \Year{1996} \Page{8591}.


\bibitem{abrikosov}
\Name{Abrikosov A. A. \And L. P. Gor'kov L. P.}
\Review{Sov. Phys. JETP} \Vol{12} \Year{1961} \Page{1243}.

\bibitem{bill}
\Name{Bill A., Kresin V. Z., \And Wolf S. A.}
\Review{Z. Phys. B} \Vol{104} \Year{1997} \Page{759}.

\bibitem{umma}
\Name{Ummarino G. A. \And Gonnelli R. S.}
\Review{Phys. Rev. B} \Vol{56} \Year{1997} \Page{R14 279}.

\bibitem{brawner}
\Name{Brawner D. A., Mamser C., \And Ott H. R.}
\Review{Phys. Rev. B} \Vol{55} \Year{1997} \Page{2788}.


\end{thebibliography}
